\documentclass[12pt,sort&compress]{elsarticle}
\usepackage{amsmath,amssymb,amsfonts}
\usepackage{algorithmic}
\usepackage{graphicx}
\usepackage{textcomp}
\usepackage{graphicx}
\usepackage{epstopdf}
\usepackage{lineno,hyperref}
\usepackage{amsmath,amssymb,amsthm}
\usepackage{mathrsfs}
\usepackage{times}
\usepackage{hypernat}
\usepackage{caption2}
\usepackage{mathrsfs}
\usepackage{times}
\usepackage{hypernat}
\usepackage{graphicx}
\usepackage{overpic}
\usepackage{color}
\usepackage{multirow}

\usepackage{lineno,hyperref}
\usepackage{float}
\journal{Elsevier}
\modulolinenumbers[200]
\begin{document}
	
\begin{frontmatter}
\title{LPIPS-AttnWav2Lip: Generic Audio-Driven lip synchronization for Talking Head Generation in the Wild\tnoteref{acceptednote}} 

\author[mymainaddress]{Zhipeng Chen }
\ead{m202110584@xs.ustb.edu.cn}
\author[mymainaddress]{Xinheng Wang }
\ead{D202210370@xs.ustb.edu.cn}
\author[mymainaddress]{Lun Xie\corref{mycorrespondingauthor}}
\ead{xielun@ustb.edu.cn}
\author[mysecondaryaddress]{Haijie Yuan\corref{mycorrespondingauthor}}
\ead{service@poteit.com}
\author[mythirdaryaddress]{Hang Pan }
\ead{12023855@czc.edu.cn}   

\address[mymainaddress]{University of Science and Technology Beijing,\\ Beijing, 100083, China}
\address[mysecondaryaddress]{Xiaoduo Intelligent Technology (Beijing) Co., Ltd,\\ Beijing, 100094, China}
\address[mythirdaryaddress]{Department of Computer Science, Changzhi University,\\Changzhi, 046011, China}

\cortext[mycorrespondingauthor]{Corresponding author}
\tnotetext[acceptednote]{This paper has been accepted by Elsevier's \textit{Speech Communication} journal. Official publication link: \url{https://doi.org/10.1016/j.specom.2023.103028}}

\begin{abstract}
\par Researchers have shown a growing interest in Audio-driven Talking Head Generation. The primary challenge in talking head generation is achieving audio-visual coherence between the lips and the audio, known as lip synchronization. This paper proposes a generic method, LPIPS-AttnWav2Lip, for reconstructing face images of any speaker based on audio. We used the U-Net architecture based on residual CBAM to better encode and fuse audio and visual modal information. Additionally, the semantic alignment module extends the receptive field of the generator network to obtain the spatial and channel information of the visual features efficiently; and match statistical information of visual features with audio latent vector to achieve the adjustment and injection of the audio content information to the visual information. To achieve exact lip synchronization and to generate realistic high-quality images, our approach adopts LPIPS Loss, which simulates human judgment of image quality and reduces instability possibility during the training process. The proposed method achieves outstanding performance in terms of lip synchronization accuracy and visual quality as demonstrated by subjective and objective evaluation results. The code for the paper is available at the following link: \url{https://github.com/FelixChan9527/LPIPS-AttnWav2Lip}
\end{abstract}

\begin{keyword}
 Audio-driven Generation, lip synthesis, LPIPS Loss, Multimodal Fusion, Talking Head Generation.
\end{keyword}

\end{frontmatter}

\linenumbers

\section{Introduction}
\par In recent times, there has been significant interest in scientific research around learning the relationship between speech and facial images. 
The primary goal is to generate realistic facial animations based on audio features. 
This technique has already found diverse real-life applications, including the creation of virtual digital humans, video dubbing, and telepresence for role-playing video games.
\par The primary challenge of Talking Head Generation lies in addressing the lip synchronization problem, which involves ensuring that the character's lips match the speech content of the given audio. 
Achieving perfect lip synchronization means automatically generating lip animation that precisely corresponds to speech content of given audios. 
This is particularly important in scenarios such as advertising, where the characters in a video need to say specific words that might not match their original lip movements. 
Similarly, in post-production manual dubbing of movies, overlaying the dubbed audio onto the original video can cause the actor's lip movements to become out of sync, negatively affecting the viewer's experience.
\par Advancements in deep neural networks have opened doors to audio driven synthesize non-existent faces or manipulate real faces in images or videos using deep learning. 
Chung et al.~\cite{ref5} introduced SyncNet, the first end-to-end lip synchronization network. 
It predicts the offset of video frames and audio without the assistance of additional audio-video encoder. 
Additionally, Chung et al. proposed a synchronization evaluation metric, Lip-Reading Similarity Distance (LRSD), to objectively evaluate the degree of semantic similarity of lip synchronization. 
Kumar et al. utilized the success of Pix2Pix~\cite{ref6} in image-to-image translation to synthesize lip synchronization from text and proposed ObamaNet~\cite{ref7}. 
However, the application of ObamaNet is very limited, only applicable to synthesize the lip animation of Obama's speech, and cannot be widely applied to other speakers. 
Vougioukas et al.~\cite{ref8} proposed an end-to-end speech-driven facial animation synchronization model based on Temporal GAN and multi-discriminator design. 
This method comprises a frame discriminator to ensure the generated face images are clear and a sequence discriminator to generate natural facial motion based on given audios. 
Zhou et al.~\cite{ref9}, inspired by Vougioukas et al.~\cite{ref8}, achieved face generation for arbitrary subjects by learning separate audiovisual representations, ignoring the correlation between head pose and audio. 
Another work by Chung et al.~\cite{ref10} (Speech2Vid) generates speaker-identity independent video animations by decoupling speaker voice and identity, but it fails to consider the continuity of the time series and hence suffers from frame skipping or jittering. 
On the other hand, Chen et al.~\cite{ref11} aim to generate face videos with accurate synchronization and clear picture quality for different face shapes, viewing angles, facial features, and noisy audio conditions. 
In contrast, Yang et al.~\cite{ref12} generate expressive talking head videos by predicting facial landmarks that reflect the dynamics of the speaker's consciousness, rather than attempting to learn a direct mapping from audio to raw pixels.
\par Current research focuses on accurately audio--driven lip movements in static images or videos of specific individuals during training, but cannot accurately modify lip movements for arbitrary identities in dynamic, unconstrained videos of speakers’ faces~\cite{ref13}. Wav2Lip~\cite{ref13} is the first speaker-independent lip synchronization framework that can synchronize the facial videos of arbitrary speakers with arbitrary speech to achieve accurate lip synchronization. It is an extension of LipGAN~\cite{ref14} and uses images of people with masked lower faces as target faces. Wav2Lip takes into account the temporal contextual relationship between frames and randomly selects five consecutive frames (T=5) as input to ensure continuity of action. Additionally, Wav2Lip uses a pre-trained Lip-Sync expert discriminator based on the SyncNet~\cite{ref5} to detect whether the lips are synchronized or not. As the reconstruction loss of the lip region only accounts for less than 4$\%$ of the total reconstruction loss, Wang et al.~\cite{ref15} incorporated an attention mechanism into the depth network to help it focus more on the lip region, which further improves the accuracy of lip synchronization.
\par However, because the encoded visual and audio features are directly concatenated as the input to the decoder and the generalization of tones is enhanced by numerous speakers' video datasets (e.g., LRS2~\cite{ref35}), the neural network is unable to learn the deep association between audio content information and mouth texture information~\cite{ref38}. Although the introduction of U-Net~\cite{ref34} architecture can improve the accuracy of face generation to some extent, the audio information will gradually decrease with the depth of the image decoder, and even the problem of audio encoder failure to work. Therefore, it is still difficult for Wav2Lip~\cite{ref13} or AttnWav2Lip~\cite{ref15} in terms of audio-driven generation of high-frequency details.
\par
\begin{figure}[H]
 \centering\includegraphics[width=15cm,height=8cm]{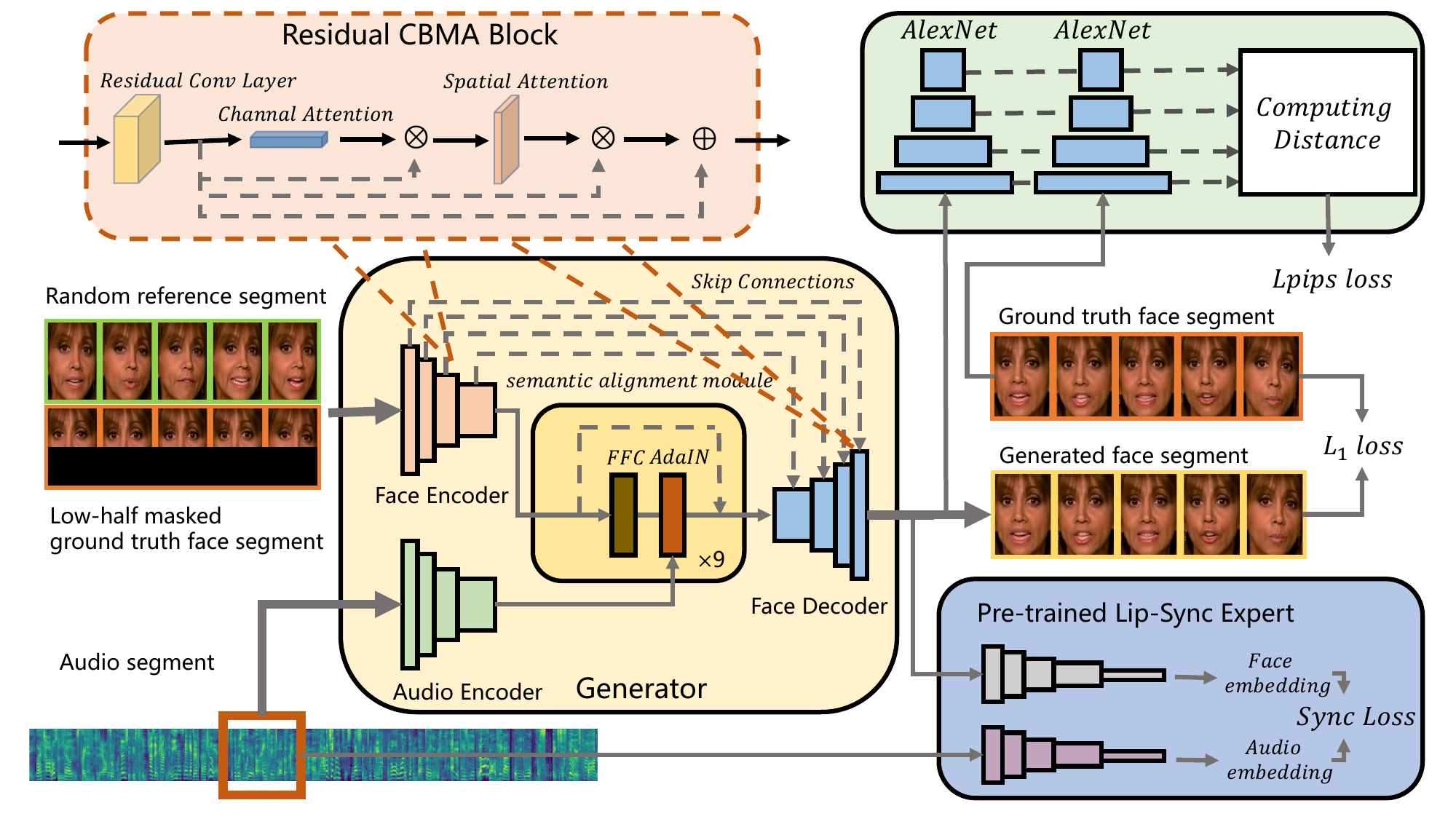}
 \caption{The framework of LPIPS-AttnWav2Lip.}\label{tu1}
\end{figure}

 Drawing inspiration from previous methods that rely on inpainting for generating talking head videos~\cite{ref13}, this paper presents a novel approach, LPIPS-AttnWav2Lip, for audio driven talking head generation. Our method, outlined in Figure \ref{tu1}, can be broken down into three main components. 
\par For the generator design, we adopt a Convolutional Block Attention Module (CBAM) based on the residual structure, inspired by Wang et al.~\cite{ref15}. CBAM improves the focus on encoding and decoding of lip region information in the channel and spatial dimensions, and suppresses the influence of irrelevant information. The generator, while using the U-Net architecture, cuts down the number of layers of encoders and decoders to alleviate the problem of audio information having less and less influence on image generation due to decoder depth on the one hand, and to reduce the training difficulty as well as to save resources on the other. 
\par Moreover, we introduce a semantic alignment module to effectively fuse audio and video features. Specifically, we expand the receptive field of the network layer using an FFC layer based on a double-branching structure to acquire local features and global contextual information about the visual feature map. We also introduce Adaptive Instance Normalization (AdaIN) to align statistical information of visual features with audio latent vector, enhancing the audio drive of the pixels in the lip region. This approach enables the alignment and adaptation of visual and audio features on the semantic information of the audio content, and does not increase the computational cost.
\par Finally, we use LPIPS loss instead of the adversarial loss in the visual quality discriminator to address training difficulties and gradient disappearance/explosion in GANs. LPIPS loss can simulate human judgments on image quality, ensuring a steady decline in image quality loss during training and reducing problems such as gradient disappearance or explosion. This approach provides a better training environment for the implementation of lip synchronization.
\par In summary, the innovations of this article can be summarized as follows:
\begin{itemize}
\item This paper introduces a novel approach to generating speaker-face videos that are accurately lip-synced with the corresponding audio. The proposed method is audio-driven and can be applied to any speaker, making it a generic solution.
\end{itemize}
\begin{itemize}
\item This paper introduces a method that employs residual CBAM blocks to improve the accuracy of lip synchronization by helping the model to reconstruct the lip region. Additionally, the semantic alignment module expand the receptive field of the network, allowing for alignment of statistical information on visual and audio features. Lastly, we analyzed the training process of GAN to find out the reasons affecting the lip synchronization effect. In order to balance the image quality and the accuracy of lip synchronization, the adversarial loss is replaced with LPIPS loss.
\end{itemize}
\begin{itemize}
\item In this paper, we evaluate the effectiveness and superiority of our proposed method by analyzing lip synchronization metrics such as LSE-C and LSE-D, as well as visual metrics such as FID.
\end{itemize}
\par The second chapter of this paper provides a review of the development of talking head generation. In the third section, the specific details of the proposed method in this study are discussed. The fourth chapter presents the training process and the dataset used in the model. The fifth section provides an analysis of the experimental results. Finally, the sixth chapter offers an outlook on future research directions.
\section{Related work}
\subsection{Talking Head generation}
\par The rapid advancement of deep learning technology and the extensive adoption of Talking Head technology has led to the flourishing of Talking Head generation methods. This paper categorizes the existing modeling methods into two groups based on their modeling approaches: 3D and 2D.
\par In 3D-based methods, typically, a population-specific 3D model needs to be constructed beforehand, and the final 3D effect is obtained through neural rendering~\cite{ref2}. Techniques such as 3D mesh~\cite{ref16} and Blendshape~\cite{ref17} have also been utilized to construct population-specific 3D models. Hussen Abdelaziz et al.~\cite{ref43} froze the pre-trained acoustic model to provide more audio prior information, and than trained the last regression layer to predict blendshape coefficients. Recently, NeRF-based methods have provided new possibilities for 3D Talking Head generation. For example, Guo et al.~\cite{ref18} learn the audio signal features via a conditional implicit function to create a dynamic NeRF and ultimately synthesize a high-fidelity video of the speaker matching the audio signal.
\par The aforementioned 3D models are costly and mostly dependent on the training of a specific identity, making it difficult to replace with a new identity. In contrast, the 2D-based approach offers a more feasible alternative. Chen et al. utilized facial landmarks as an intermediate process to map audio signals to video frames~\cite{ref11}. Additionally, GAN-based end-to-end approaches have become the mainstream idea. Agarwal et al.~\cite{ref44} proposed AVFR-GAN to animate a source image by transferring head motion from a driving video using a dense motion field generated using learnable keypoints. However, the method based on driving a single image through a video or audio is very limited in application and often distort the background. Venkatesh et al. proposed a Transformers-based visual-auditory detector to estimate the degree of synchronization between lip motion and speech~\cite{ref19}. Prajwal et al. used a pre-trained Lip-Sync expert discriminator to correct the lip shape of the generated face and a visual quality discriminator to remove artifacts in the generation process~\cite{ref13}. Meanwhile, Wang et al. placed more emphasis on the accuracy of lip synchronization and utilized an attention mechanism to guide the network towards focusing more on reconstructing the lip region~\cite{ref15}.
\subsection{Speech Content Semantic Alignment}
In order to make the final generated lip synchronized with the speech content, after encoding the audio and image, it is necessary to make the visual information semantically aligned with the target audio information in speech content by feature fusion. MakeItTalk~\cite{ref39} enables audio content to indirectly drive facial pixels by directly predicting facial keypoints from audio features. Wav2Lip~\cite{ref13} and AttnWav2Lip~\cite{ref15} combine visual and audio embedding as input to the image decoder by concatenate after encoding visual and audio information. However, this simple concatenate approach does not reflect the speech content information to the final generated face well, because there may not be correlation between visual and audio features, and the audio information may be ignored in the subsequent learning. AVFR-GAN~\cite{ref44} obtains audio-visual attention by calculating the dot product of the motion feature map and audio embedding at each location. Guan et al.~\cite{ref45} proposed StyleSync framework, which encodes face information and audio information into W+ space as input information for style-based generator. The article~\cite{ref40} disentangles the face information into Person-ID space and Word-ID space by means of adversarial learning, and then interchanges the speech content information of reference face and audio are realized. However, this method is more difficult to train and does not guarantee that the face information can be completely separated from the Person-ID space and Word-ID space. Sun et al.~\cite{ref46} proposed a three-branch convolution-Transformer-hybrid backbone network. the reference branch and the audio branch provide the information of the lower half of the face missing by the mask part and the mouth shape in the main branch.
\subsection{LPIPS Loss}
\par Richard et al.~\cite{ref26} introduced the Learned Perceptual Image Patch Similarity (LPIPS) concept to help computers in replicating human image quality discrimination. Distinguishing between two distorted images and determining which one is closer to the original is simple for humans but is challenging for computers. The LPIPS metric relies on the human visual system breaking down an image into small, perceptually significant patches and then comparing the differences of these patches between the two images. In deep learning, LPIPS is widely used as a loss function for image generation tasks like image-to-image translation, super-resolution, and style transfer. Its primary objective is to train the neural network to generate images that are perceptually similar to the target image, instead of merely matching pixel. By using the LPIPS Loss, Zheng et al.~\cite{ref27} enhanced BasicVSR~\cite{ref28} to recover compressed video textures and fulfill human perception criteria.

\section{Method}\label{sec3}
\subsection{Basic Framework}
\par The generator in our method is made up of three main components like the Wav2Lip framework ~\cite{ref13}: the audio encoder, the face encoder, and the face decoder. These three main components use a 2D-CNN with a residual structure, and the face encoder and the face decoder form the U-Net~\cite{ref34} structure. The lower half of the face is masked for ground truth as well as randomly selected consecutive reference frames (T=5) concatenated as input to the face encoder and encoded into visual feature maps. The audio encoder encodes corresponding audio segments to audio latent vector after MFCC processing. The encoded visual feature maps and audio latent vector are fused by the semantic alignment module and then passed into the face decoder to generate face images on a frame-by-frame basis. Inspired by the use of CBAM to improve lip synchronization accuracy in article~\cite{ref15}, we add residual CBMA blocks between every two convolution layers of face encoder and decoder to enhance the feature representation and improve the robustness of the model.

As for training the generator, the loss function \eqref{Equation(6)} is the weighted sum of the reconstruction loss $L_{{recon}}$ , synchronization loss $L_{{sync}}$ ~\cite{ref13}, and adversarial loss $L_{{gen}}$ .

\begin{equation}
L=(1-\alpha-\beta) \cdot L_{{recon }}+\alpha \cdot L_{sync}+\beta \cdot L_{gen}\label{Equation(6)}
\end{equation}

\par where $\alpha$ and $\beta$ are empirically set to 0.03 and 0.07 respectively.

\subsection{Audio-Visual Semantic Alignment}
\par
\begin{figure}[H]
 \centering\includegraphics[width=13cm,height=4cm]{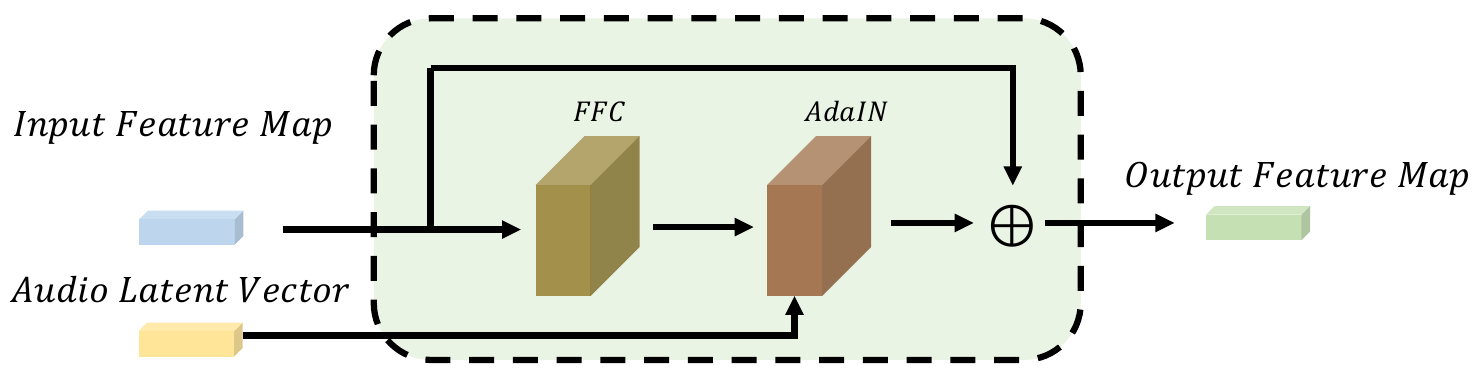}
 \caption{The specific structure of the Semantic Alignment Module.}\label{tu3}
\end{figure}

 The semantic alignment module comprises two main components: the Fast Fourier Convolution layer (FFC) and the Adaptive Instance Normalization (AdaIN), as depicted in Figure \ref{tu3}.
\par In this study, lip synchronization is accomplished using the following steps: firstly, a video frame at a specific point in time is masked, and the masked image is then merged channel-wise with a randomly selected video frame from the same video. Next, the corresponding audio segment and the randomly selected reference frame are employed as inputs to generate a face image comprising the lower half of the face. This process is essentially conditional inpainting of the masked face image~\cite{ref29}. Since the masked region takes up half of the face image, the lip synchronization task necessitates considering both the semantic content features associated with the audio and the additional information offered by the upper half of the face image. As a result, the task requires consideration of the image information from a global perspective.
\par
\begin{figure}[H]
 \centering\includegraphics[width=14cm,height=6cm]{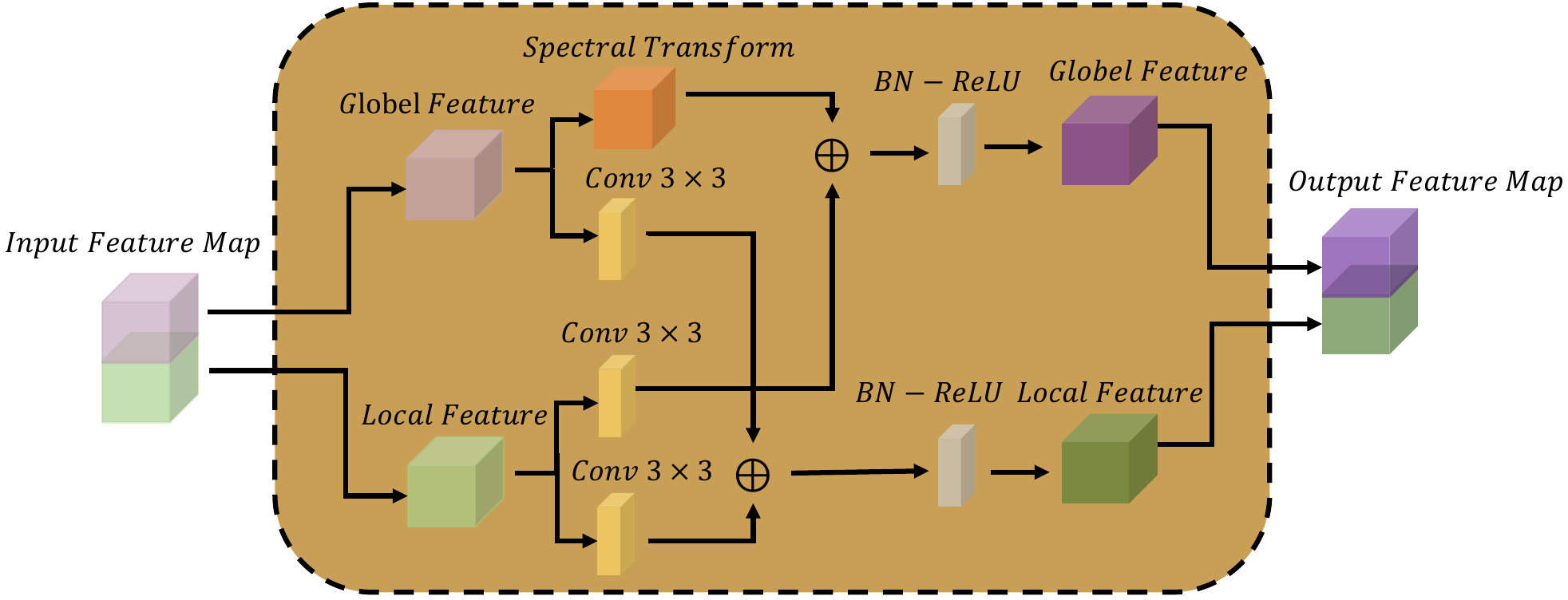}
 \caption{The specific structure of the FFC Block.}\label{tu4}
\end{figure}

 However, inpainting-based face generation methods~\cite{ref13,ref15} may encounter information leakage ~\cite{ref29} due to the mutual inpainting condition of the masked face image and the random reference face. According to the reference~\cite{ref30}, the network layer of the image encoder plays a critical role in the receptive field of the image in large masking-based image inpainting tasks. The larger the receptive field of the network layer, the more contextual information the image encoder can acquire, thereby enhancing the masking inpainting. However, conventional convolution can only capture local image information and has limitations for large mask-based image inpainting tasks. To address this problem, Lu et al.~\cite{ref31} proposed an FFC-based residual network structure that effectively restores large masked images to SOTA level, as depicted in Figure \ref{tu4}. The method uses a two-branch structure, with the local branch utilizing conventional convolution to obtain local features on the image, while the global branch obtains the global contextual information of the image based on the fast Fourier transform of the channel dimension.
\par In our work, we have incorporated the two-layer FFC residual stacking structure proposed by Suvorov et al.~\cite{ref30} to enhance the model's robustness and improve its ability to learn the global structure of the image. 
\par To achieve lip synchronization, the lip shape information of the randomly selected reference frames are crucial since it contains the original speech content semantic information from itself. However, this information needs to be transformed into the target speech content semantic information to achieve the target lip synchronization. Therefore the original changed information is not conducive to the conversion of the target information.
\par Inconsistencies in the semantic information of the lip shape across different random reference frames pose a challenge. Therefore, a learning compensation approach is necessary to address this issue~\cite{ref32}. However, using only the convolutional layer for learning can lead to training difficulties because it requires mapping the reference frame and speech spectrum into the same semantic information domain. Since the speech content semantic information of the lip shape in each random video reference frame varies from its distance in the semantic information domain for a given speech spectrum, a more efficient approach is needed to address these issues.
\par We incorporated the design principles of reference~\cite{ref29} and integrated the AdaIN algorithm~\cite{ref32} allows better semantic alignment of visual features with audio features. Specifically, we substituted the second FFC layer in the two-layer FFC residual stacking structure with the AdaIN layer. The AdaIN algorithm utilizes an adaptive affine transformation technique, which assists in effectively fusing and transforming feature information to any given feature, as shown in Equation\eqref {Equation(10)}. 
\begin{equation}\label{Equation(10)}
\operatorname{AdaIN}(x, y)=\operatorname\sigma(y)\left(\frac{x-\operatorname\mu(x)}{\operatorname\sigma(x)}\right)+\operatorname\mu(y)
\end{equation}
\par where $x$ and $y$ represent different characteristics.$\sigma$ and $\mu$ represent mean and variance calculations, respectively.
\par In our task, AdaIN aligns the mean and variance of the encoded video feature map with the encoded audio latent vector of the speech spectrum in the channel dimension after the previous layer of processing, allowing the feature map processed by AdaIN to exhibit a high mean activity similar to that of the audio latent vector at the output. This leads to better preservation of the semantic information of the audio in our task.
\subsection{Tradeoff: Lip Accuracy vs. Image Quality}
\par The training of GANs is highly complex and can result in issues such as mode collapse~\cite{ref47}, gradient disappearance or explosion. These issues can ultimately have an adverse effect on the quality of the generated output. When GANs seek a discontinuous mapping in the space of continuous mappings, they may fail to converge or converge to a singular branch, causing mode collapse~\cite{ref47}. In this situation, GANs tend to generate samples that lack diversity. To overcome these challenges, it is essential to adopt a more stable loss function, which has been proven to be an effective solution. 
\par During the early stage of GAN training, the focus of the network framework is on modeling the overall face of the person, with little emphasis on the lip shape, which accounts for only 4$\%$ of the face. At this stage, the adversarial training discriminates the generated image from the real image, leading to a significant improvement in visual quality. However, achieving a better lip synchronization is challenging and can be influenced by the training effect and method used, as seen in LipGAN~\cite{ref14}. 
\par In the later stage of training, when the character face modeling reaches a certain level, the model needs to prioritize the synchronization of audio and lip shape to improve lip synchronization. To achieve this, Wav2Lip adds SyncNet loss to penalize unsynchronized lip shape. However, before adding SyncNet loss, the image generator and discriminator are already in a stable adversarial equilibrium state. Adding SyncNet loss at this point will affect the quality of the generated images by the generator, destroying the original adversarial equilibrium state and increasing the difficulty of model training and the likelihood of gradient disappearance or gradient explosion.
\par Therefore, in this study, we use LPIPS loss instead of the adversarial loss of the visual discriminator. This more stable loss function ensures stable loss convergence of the model during training without compromising visual quality, reduces issues such as gradient disappearance or gradient explosion, and creates a favorable training environment for lip synchronization. As demonstrated by Phillip et al.~\cite{ref33}, using only L1 or L2 reconstruction loss in the loss function of the generator can increase the probability of blurry generated images. This is because L1 and L2 reconstruction loss are adequate for capturing the low-frequency features of the image data, while the discriminator only needs to capture high-frequency features such as texture details to further improve image quality. The purpose of using LPIPS loss is to emulate human perception of image quality. By using a patch-based approach to assess image quality, the generator can be effectively trained to capture high-frequency features in images. Thus, we contend that this patch-based perceptual loss function can partially substitute for the visual quality discriminator.
\par To compute the LPIPS loss, the first step is to pre-process the generated and target images. This involves extracting feature mappings at various network layers through pre-trained neural networks such as VGG or AlexNet. In this paper, we adopt the approach proposed in reference ~\cite{ref26}, which uses the AlexNet model. Further details can be found in reference~\cite{ref26}. Subsequently, the distance between the feature mappings of the two images is measured using a distance metric (L2 parametric) to obtain the LPIPS loss. The calculation of LPIPS loss is represented by Equation\eqref {Equation(11)}:
\begin{equation}\label{Equation(11)}
L_{LPIPS}=\sum_l \frac{1}{H_l W_l} \sum_{h, w}\left\|w_l \odot\left(\hat{y}_{h w}^l-\hat{y}_{0 h w}^l\right)\right\|_2^2
\end{equation}
\par Where $H_l$ and $W_l$ are the number of patches of the feature map at the layer $l$. $h$ and $w$ are the patches at the corresponding positions. $w_l$ is the weight vector used for channel domain normalization. $\hat{y}_{h w}^{l}$ and $\hat{y}_{0 h w}^{l}$ represent the ground truth feature map of the layer $l$ and the feature map of the generated image, respectively.
\par Hence, the minimization of Equation\eqref {Equation(6)} can be modified to Equation\eqref {Equation(12)}:
\begin{equation}\label{Equation(12)}
L=(1-\alpha-\beta) \cdot L_{{recon }}+\alpha \cdot L_{s y n c}+\beta \cdot L_{LPIPS}
\end{equation}

\section{Training}
\par Our method has been implemented using the PyTorch framework. Below, we provide specific training details for our approach, as well as an introduction to the corresponding dataset.
\subsection{Dataset}
\par We utilized three datasets, namely LRS2~\cite{ref35}, LRS3~\cite{ref36}, and LRW~\cite{ref37}, to evaluate the effectiveness of our proposed approach. The LRS2 dataset consists of over 1000 hours of lip synchronization videos from various accents and backgrounds sourced from BBC TV shows. On the other hand, LRS3 comprises more than 500 hours of video data obtained from several TV shows and movies. In contrast, the LRW dataset has video data sourced from social media platforms such as YouTube.
\subsection{Evaluation metrics}
\par In this study, our proposed method is evaluated based on two criteria: visual quality and lip synchronization. To evaluate the visual quality, we use the FID metric, which calculates the Fréchet distance between the real image set and the generated image set. A lower FID value indicates a more similar distribution of the generated and real images, and thus a better generation quality. As for the lip synchronization accuracy, we refer to Prajwa et al.'s ~\cite{ref13} LSE-C (lip sync Error - confidence) and LSE-D (lip sync Error - distance) metrics. A higher LSE-C value indicates a greater audio-video correlation, while a lower LSE-D value indicates a more pronounced lip synchronization accuracy.
\subsection{Training details}
\par The parameter settings in this study are in line with the works of Prajwal~\cite{ref13} and Wang~\cite{ref15}. The only difference is that in this study, the size of the feature map obtained after concatenating the ground truth and random reference frames that cover the lower half of the face through the face encoder is $6\times {6}\times {256}$. Similarly, the audio clip passed through the audio encoder also obtains a feature map of the same size. The semanstic alignment module comprises one FFC layer and one AdaIN layer. The first semanstic alignment module takes the output of the face encoder, and the subsequent semanstic alignment module takes the output of the previous one. Since the feature fusion process consists of nine semanstic alignment modules, the audio features are input nine times.  The output size of the last semanstic alignment module is $6\times {6}\times {256}$, which serves as the input of the face decoder. The face decoder generates a $96\times {96}$ size face image. By introducing the LPIPS loss, the training process becomes stable, and the training time is reduced to less than 48 hours.
\section{Experiment}
\subsection{Comparison with state-of-the-art Methods}
\par We have evaluated the proposed method by comparing it to four well-known 2D-based talking head generation methods, namely LipGAN~\cite{ref14}, Speech2Vid~\cite{ref10}, Wav2Lip~\cite{ref13}, and AttnWav2Lip~\cite{ref15}, using the same experimental setup. Wav2Lip adopts the network structure of LipGAN but is different in that it includes both Wav2Lip, which only uses a pre-trained Lip-Sync expert discriminator, and Wav2Lip-GAN, which includes both a Lip-Sync expert discriminator and a visual quality discriminator. AttnWav2Lip, on the other hand, introduces a CBAM attention mechanism in Wav2Lip-GAN to improve lip synchronization by allowing the model to focus more on the reconstruction of the lip region. Speech2Vid, proposed by Chung et al.~\cite{ref10}, is an encoder-decoder structure-based lip-sync model. We present the results of comparing our method with these four state-of-the-art approaches in Table \ref{table1}, in terms of both visual quality and lip synchronization.
\par Table \ref{table1} shows the comparison of visual quality and lip synchronization performance of our proposed method with four other 2D-based talking head generation approaches. Among them, Wav2Lip-GAN with a visual quality discriminator achieves the best performance in terms of the visual quality metric FID. Our proposed method exhibits the second-best FID metric and performs comparably to Wav2Lip-GAN on the LRS3 and LRW datasets. Conversely, Wav2Lip without a visual quality discriminator shows the worst performance. Our proposed method achieves the best lip synchronization performance on all three lip synchronization datasets, as indicated by the lip synchronization metrics LSE-C and LSE-D. The lowest LSE-D value implies that our method generates the closest distance between lips and audio representation, the smallest temporal deviation between speech and lips, and the best lip synchronization performance. The highest LSE-C value suggests that our proposed method has the highest average confidence level and the closest correlation between video-audio, resulting in the most comfortable visual sensation for lip synchronization.
\par
\begin{figure}[H]
 \centering\includegraphics[width=16cm,height=9cm]{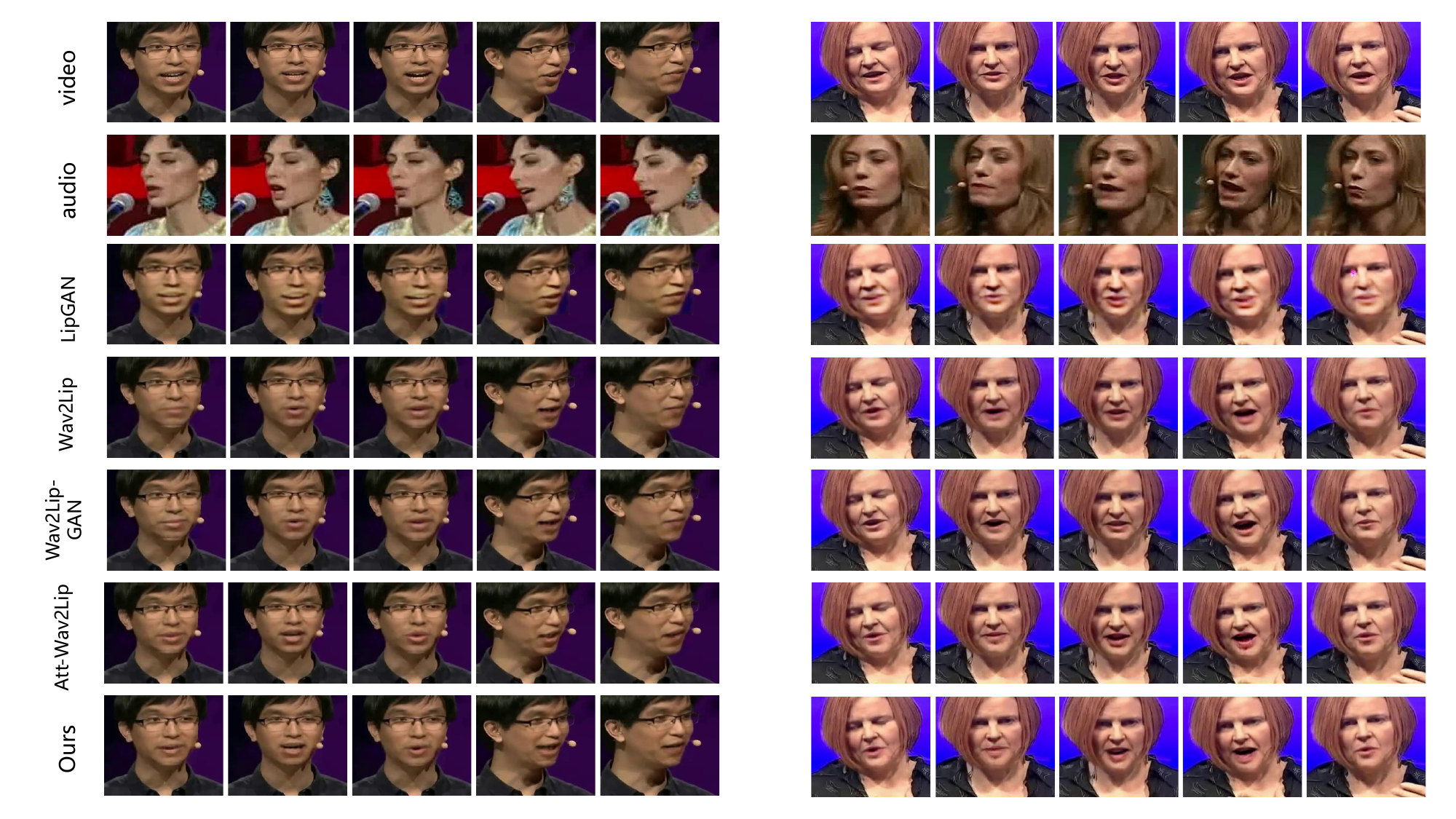}
 \caption{We conducted a visual comparison between our proposed method and several representative 2D-based talking head generation methods, including Wav2Lip/Wav2Lip-GAN~\cite{ref13}, AttnWav2Lip\cite{ref15}, and LipGAN\cite{ref14}.}\label{tu5}
\end{figure}

 Our proposed model has demonstrated superior performance in terms of lip synchronization and visual quality. We believe that the LPIPS loss effectively simulates human judgment of image quality and compensates for the deficiency of the visual quality discriminator in improving the generated image quality. The use of LPIPS loss also stabilizes the training process and mitigates issues such as training difficulties and gradient explosion. Additionally, CBAM improves the model's attention to the lip region and enhances lip synchronization. The semanstic alignment module facilitates the integration of audio and video features, while the two-branch FFC layer captures both local and global contextual information of the image. Furthermore, AdaIN preserves the semantic information in audio better. Our approach outperforms the classical approaches presented in the reference~\cite{ref10,ref13,ref14,ref15} in terms of both image quality and audio-visual synchronization, as demonstrated in Figure \ref{tu5} and the accompanying Table \ref{table1}.
\begin{table}
\centering
\caption{We conducted a comparison of the LRS2, LRS3, and LRW datasets with five representative methods. The evaluation criteria were based on the visual quality FID metric and the lip synchronization metrics LSE-C and LSE-D. The best-performing method is highlighted in bold.}
\label{table1}
\resizebox{\textwidth}{!}{
\begin{tabular}{cllcllcllc}
\hline
\multirow{2}{*}{METHOD} & \multicolumn{3}{c}{LRS2~\cite{ref35}}                                                                     & \multicolumn{3}{c}{LRS3~\cite{ref36}}                                                                     & \multicolumn{3}{c}{LRW~\cite{ref37}}                                                                      \\ \cline{2-10} 
                        & \multicolumn{1}{c}{LSE-C↑} & \multicolumn{1}{c}{LSR-D↓} & FID↓                               & \multicolumn{1}{c}{LSE-C↑} & \multicolumn{1}{c}{LSR-D↓} & FID↓                               & \multicolumn{1}{c}{LSE-C↑} & \multicolumn{1}{c}{LSR-D↓} & FID↓                               \\ \hline
Wav2Lip~\cite{ref13}         & 6.761                      & 7.151                      & \multicolumn{1}{l}{4.922}          & 7.146                      & 7.002                      & \multicolumn{1}{l}{4.682}          & 6.532                      & 7.185                      & \multicolumn{1}{l}{3.186}          \\
Wav2Lip-GAN~\cite{ref13}     & 6.782                      & 7.208                      & \multicolumn{1}{l}{\textbf{4.469}} & 7.067                      & 7.096                      & \multicolumn{1}{l}{\textbf{4.054}} & 6.412                      & 7.341                      & \multicolumn{1}{l}{\textbf{2.457}} \\
AttnWav2Lip~\cite{ref15}    & 6.834                      & 7.171                      & \multicolumn{1}{l}{4.979}          & 7.086                      & 7.071                      & \multicolumn{1}{l}{4.614}          & 6.581                      & 7.143                      & \multicolumn{1}{l}{3.156}          \\
Ours                    & \textbf{7.287}             & \textbf{6.834}             & 4.624                              & \textbf{7.513}             & \textbf{6.706}             & 4.563                              & \textbf{6.86}              & \textbf{6.989}             & 2.685                              \\
LipGAN~\cite{ref14}          & 3.199                      & 10.33                      & -                                  & 3.193                      & 10.65                      & -                                  & 3.35                       & 10.05                      & -                                  \\
Speech2Vid~\cite{ref10}    & 1.587                      & 14.23                      & -                                  & 1.681                      & 13.97                      & -                                  & 1.762                      & 13.14                      & -                                  \\ \hline
\end{tabular}
}
\end{table}
\subsection{Ablation experiments.}
\par The ablation experiments in this paper contain two parts that explore the effects of the semantic alignment module and the LPIPS loss on lip synchronization, respectively. Where CBAM+Align indicates that our model uses CBAM and the semantic alignment module (i.e., the LPIPS loss is removed in our method).  CBAM+LPIPS indicates the use of CBAM and LPIPS loss (i.e., the semanstic alignment module is removed in our method). 
\par According to the results in Table \ref{table3}, we find that the semantic alignment module can significantly improve the lip synchronization. This also shows that using a simple concatenate of audio features and visual features does not allow visual features to be well associated with audio features in a deep way. It is difficult to approach the target audio semantic information by concatenate method because of the various speech content semantic information of the reference frames itselves. And because of the deep U-Net architecture used in Wav2Lip and AttnWav2Lip, the face encoder compresses the visual information to a one-dimensional vector size before inputting it into the decoder. Therefore, in the subsequent decoding process, the audio semantic information will have less and less influence on the final lip generation, and even the influence of audio information will be ignored. We make the U-Net architecture shallower to make audio semantic information more likely to survive in the decoding stage, as shown by the comparison between CBAM+Align$\times 1$ and AttnWav2Lip. When we stack the semantic alignment modules by cascading them 9 times (CBAM+Align$\times 9$) for a more adequate semantic alignment of visual and audio features. It can be found that CBAM+Align$\times 9$ will be better than CBAM+Align$\times 1$ in terms of lip synchronization accuracy. This shows that a more adequate semantic alignment of visual and audio features will allow more audio semantic information to be injected into the visual information, thus making the lip region pixel probability distribution more consistent with the distribution of audio semantic information. 
\par As we expected, the results presented in Table \ref{table4} show that the introduction of LPIPS loss can improve the lip synchronization accuracy. As mentioned in Section \ref{sec3}, stable loss is crucial for lip synchronization videos. A stable training environment ensures that the network accurately captures the mapping relationship between audio and lip image, resulting in a realistic and exact lip synchronization. This effect not only enhances the visualization and impact of the video, but also improves the user's viewing experience and engagement. 
\par The specific training processes are implemented on the LRS2 dataset.
\begin{table}
\centering
\caption{Ablation experiment 1. The aim is to investigate the effect of the semantic alignment module on lip synchronization accuracy.}
\label{table3}
\begin{tabular}{ccc}
\hline
Method               & LSE-C↑         & LSR-D↓         \\ \hline
CBAM+Align$\times 9$             & 6.902          & 7.083          \\ \hline
CBAM+Align$\times 1$             & 6.837          & 7.092          \\ \hline
AttnWav2Lip~\cite{ref15} & 6.834          & 7.171          \\ \hline
\textbf{Ours}        & \textbf{7.287} & \textbf{6.834} \\ \hline
\end{tabular}
\end{table}
\begin{table}
\centering
\caption{Ablation experiment 2. The aim is to investigate the effect of LPIPS loss on lip synchronization accuracy.}
\label{table4}
\begin{tabular}{ccc}
\hline
Method               & LSE-C↑         & LSR-D↓         \\ \hline
CBAM +LPIPS          & 7.093          & 6.972          \\ \hline
AttnWav2Lip~\cite{ref15} & 6.834          & 7.171          \\ \hline
\textbf{Ours}        & \textbf{7.287} & \textbf{6.834} \\ \hline
\end{tabular}
\end{table}
\subsection{Subjective Judgment}
\par Manual evaluation is crucial in assessing lip synchronization in addition to objective metrics. Therefore, we designed a user questionnaire to further evaluate our proposed method. To perform this evaluation, we generated 12 talking videos using our method on the LRS2 dataset from various Audio-Video sources and compared them with videos generated by two state-of-the-art methods (Wav2Lip-GAN and AttnWav2Lip). The questionnaire consisted of five different scores, ranging from 1-5, where higher scores indicate higher participant satisfaction levels. We collected 384 survey comments from a total of 32 participants. Table \ref{table2} shows the participant's satisfaction level with the generated videos in terms of lip synchronization and visual quality. The results show that our proposed method received a high rating in terms of lip synchronization sensory aspects, and surprisingly, a significant number of participants also gave high approval to the visual quality.
\begin{table}
\centering
\caption{Experimental findings of subjective judgment.}
\label{table2}
\begin{tabular}{ccc}
\hline
Methed               & Visual Quality↑ & Lip-sync Quality↑ \\ \hline
Wav2Lip-GAN~\cite{ref13}  & \textbf{3.781}  & 2.197             \\ \hline
AttnWav2Lip~\cite{ref15} & 2.125           & 3.208             \\ \hline
Ours                 & 3.333           & \textbf{3.771}    \\ \hline
\end{tabular}
\end{table}
\section{Conclusion and Future Works}
\par This paper presents LPIPS-Wav2Lip, a novel method for genericl audio-driven talking head generation. Our approach comprises three key components: First, the model employs residual CBAM to identify and emphasize important regions and content in the input. Second, a cascaded semanstic alignment module facilitates high-quality transform and fusion between audio and video features. Finally, we train the network steadily using LPIPS loss, resulting in realistic and exact lip synchronization.
\par During the questionnaire survey, many participants expressed surprise at the lip synchronization produced by our method, but also felt that the lower video resolution reduced their viewing experience. This is due to two factors. Firstly, the training dataset is low-resolution, and secondly, our reference Wav2Lip does not consider the need for sharper face images, resulting in an output face image size of only $96\times 96$. Therefore, we plan to conduct further research and analysis in generating higher resolution talking face videos in our future work. Our aim is to produce HD talking face videos with improved perceptual effects, providing users with a more immersive and engaging experience.

\section*{CRediT authorship contribution statement}
\par Zhipeng Chen and Xinheng Wang: Conceptualization, Methodology, Visualization, Software, Writing. Lun Xie and Haijie Yuan: Supervision. Hang Pan: Data curation.
\section*{Declaration of Competing Interest}
\par The authors declare that there is no conflict of interest regarding the publication of this paper.
\section*{Acknowledgments}
\par  This work was supported in part by the Beijing Natural Science Foundation under Grant L192005.
\section*{Data availability}
\par  Data will be made available on request.

\end{document}